\documentclass[aps,prl,twocolumn,groupedaddress,showpacs,amsfonts]{revtex4}

\usepackage{graphicx}
\begin{document}


\title{Induced paramagnetic states by localized $\pi $-loops in grain boundaries}


\author{Giacomo Rotoli}
\email[]{rotoli@ing.univaq.it}
\homepage[]{http://ing.univaq.it/energeti/research/Fisica/supgru.htm}
\affiliation{Dipartimento di Energetica and Unit\'{a} di Ricerca INFM,\\ 
Universit\'{a} di L'Aquila, ITALY}


\date{\today}

\begin{abstract}
Recent experiments on high-temperature superconductors show paramagnetic behavior localized at grain boundaries (GB). This paramagnetism can be attributed to the presence unconventional $d$-wave induced $\pi$-junctions. By modeling the GB as an array of $\pi$ and conventional Josephson junction we determine the conditions of the occurrence of the paramagnetic behavior. 
\end{abstract}


\maketitle

The discovery of spontaneous currents in granular high-T$_c$ superconductors \cite{KirtleyYBCO,Mannhart,KirtleyTsuei,Tafuri,floriana} was a strong indication that a $d$-wave symmetry of the order parameter is present in these materials. Indeed the $d$-wave scenario implies the possibility of existence of so-called $\pi$-junctions, i.e., Josephson junction formed between superconductors with unconventional pairing which cause a $\pi$ shift in the phase-current relation \cite{VanH}. A $\pi$-loop is an unconventional superconducting loop which contains an odd number of $\pi$-junctions. In zero field the ground state of a $\pi$-loop shows two energy degenerate magnetization states corresponding to two spontaneous current states, clockwise and counterclockwise. In non-zero magnetic field these spontaneous currents would act like orbital currents in paramagnetism \cite{SigrRice1}. Therefore, if the sample were field cooled, to permit the inner loops to feel the magnetic field, the response would be paramagnetic as indeed it was found in early works on Paramagnetic Meissner Effect (PME) in BSSCO by Braunisch et al. \cite{Braunisch}. PME was observed also in different high-$T_c$ ceramic materials \cite{KirtleyMota}. 

However, the presence of PME in conventional low-$T_c$ samples \cite{Nbdisks} shows that it cannot always be attributed to $d$-wave pairing \cite{GFS}. Recently experiments and simulations were devised to test the relation between multiple-connectiveness and PME in conventional systems. A square array of LTC junctions was field cooled and shown to be paramagnetic over a large interval of magnetic field \cite{Moreira,Nielsen,Deleo}. These papers also proposed a qualitative explanation for the effect based on the array multiple-connectiveness rather than the presence of $\pi$-junction. The effect of adding $\pi$-junction in square arrays was analyzed in \cite{letter}.

The observation of spontaneous currents in YBCO biepitaxial 0$^\circ$-90$^\circ$ tilt-tilt
and twist-tilt grain boundaries (GB) junctions \cite{floriana} indicates that paramagnetic effects due to $d$-wave pairing could be observed in GBs. In addition, in \cite{Tafuri} spontaneous magnetic moments have been observed both in the High-T$_c$ films, where granularity or defects pin some vortices, and along the GB. Nevertheless, the sample response in field cooling was diamagnetic. A recent experiment by E. Il'ichev et al. \cite{TafuriIli} found that YBCO biepitaxial 45$^\circ$ asymmetric GB junctions in (nominally) zero field cooling show a paramagnetic response at low field. The origin of this paramagnetism could be debated. Is this simply due to presence of localized $\pi$-loops that will act similarly to two-dimensional systems \cite{letter} or can be explained by means of paramagnetic quasi-particle currents due to existence of midgap states \cite{Higashitani} ? Here we want explore in detail the first alternative. 

\begin{figure}
\includegraphics{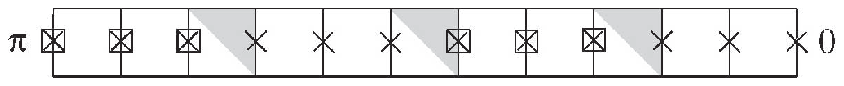}
\caption{Mixed $\pi$/conventional one-dimensional Josephson junction array with localized $\pi$-loops (half-gray).}
\end{figure}

In general a loop containing $p$ Josephson junctions will have different magnetization states when a magnetic field is applied. If the junctions are identical the loop current $I_n$ is a solution of the following equation \cite{icem2002}: 
\begin{equation}
\frac{I_n}{I_0}=\sin \left( \frac{1}{p}\left( 2\pi n-k\pi -2\pi f-\beta\frac{I_n}{I_0}\right)
\right)
\end{equation}
where $n=0,1,..,p-1$ is the quantum number in the flux quantization expression. $f$ is the frustation equal to the external flux normalized to flux quantum $\Phi_{0}$ and $\beta$ is the SQUID parameter $2\pi I_{0}L/\Phi_{0}$ with $L$ the loop inductance and $I_0$ the critical current of junctions in the loop. Varying $n$ gives different families of independent solutions within a $2\pi$ phase change \cite{notameta}. $k$ is an index which is equal to $1$ if there are an odd number of $\pi$-junctions in the loop and equal to zero otherwise.  

For any $p$ the lowest energy solutions of Eq.(1) are diamagnetic for conventional loops and paramagnetic for $\pi$-loops \cite{icem2002}. When $\beta<1$ we have only one solution in the $p=1$ loop which is diamagnetic in conventional loop and paramagnetic in $\pi$-loop without spontaneous currents \cite{KirtleyTsuei}. But in multi-junctions loops ($p>1$) we can have more states due to the presence of non-trivial solutions when changing quantum number $n$. This implies that $\pi$-loops with, e.g., $p=2$ will show spontaneous currents also for low $\beta$'s. Indeed for small $\beta$ the solutions of Eq.(1) can be written as: $\gamma _{\pm }\simeq \sin \left( \pm \pi /2-\pi f\right) \left( 1-\cos \left( \pm \pi /2-\pi f\right) \beta _{L}/2\right)$. So for $f=0$ we have two opposite spontaneous currents. For $0<f<1/2$ the solution $\gamma _{+}$ is positive (paramagnetic)\ and $\gamma _{-}$ is negative (diamagnetic). Moreover $\gamma _{+}<\gamma _{-}$ giving a lower energy for the paramagnetic solution \cite{Rio2003}. 

Small $\beta$ $\pi$-loops could be likely localized between GBs with different orientation along a junction \cite{Walker} or where faceting cause an imperfect not completely flat GB passing from a conventional junction to a $\pi$-junction or
viceversa. Recently also engineered "zigzag" arrays of mixed $\pi$/conventional junctions have been realized and measured \cite{Smilde,Hilgen03}. These can be described as an array of $\pi$-loops separated by all conventional or all $\pi$ regions \cite{Edward}.

In the following we will describe the GB as an 1d array of $N+1$ Josephson junctions placed along it. The $\pi$ additional phase is supposed to vary along the array giving arise to $\pi$ and conventional sections separated by localized $\pi$-loops (see Fig.1) 
\cite{KirtleyMoler,Mints}. We assume that system is not disordered. The magnetization dynamics of this $N$-loops system can be described using the Discrete Sine-Gordon equation (DSG) \cite{Studies}:
\begin{widetext} 
\begin{equation}
\varphi _{j,tt}+\alpha \varphi _{j,t}+\left( -1\right) ^{k\left( j\right)
}\sin \varphi _{j}=\frac{1}{\beta}\left( \varphi _{j+1}-2\varphi
_{j}+\varphi _{j-1}\right) +\frac{2\pi }{\beta}\left(
f_{j^+}-f_{j^-}\right)
\end{equation}
\end{widetext}
where $\varphi _{j}$ is the phase of the $j$-th junction in the GB, $f_{j^{\pm}}=\frac{\Phi_{ext,j^{\pm}}}{\Phi_0}$ is the frustation in the $j^{\pm}$-th loop preceding (-) or following (+) the $j$-th junction; the index $k(j)$ will be $0$ for conventional junctions and $1$ for $\pi$ junctions. Times are normalized with respect to Josephson plasma frequency $\omega_J$ and $\alpha$ is the normalized conductance. To include boundaries we set $\varphi_0=\varphi_1$, $\varphi_{N+1}=\varphi_{N+2}$ and $f_0=f_{N+1}=0$. We assume $f_{j}$ constant equal to $f$ for $1<j<N$. This implies that the magnetic field enters as boundary conditions on the two side loops of the array. The term $\frac{2\pi f}{\beta ^{1/2}}$ is equal to the normalized magnetic field at boundary: $\eta=\frac{2\pi }{\Phi _{0}}\cdot \lambda_L\lambda_J B_{ext}$ (see Ref.\cite{Studies}).  Eq.(2) is analogous to that deduced in the continuous limit by E. Goldobin et al. in \cite{Edward} in the context of analysis of "zigzag" arrays. We note that can be shown that Eq.(2) for $N$ equal one implies Eq.(1) for $p$ equal two. 

In YBCO GB junctions the Josephson length $\lambda_J$ is smaller than GB physical dimension $L$, thus the normalized length $l=L/\lambda_J$ is larger than one \cite{Tafuri}. The grain dimension along the GB $\Delta x$ is usually smaller than $L$, being roughly of $1$ $\mu$m for GB of Ref.\cite{TafuriIli} or also less in other circumstances \cite{Tafuripersonal}. GBs faceting is even smaller, ranging around $0.1-0.01$ $\mu$m \cite{Mannhart,TafuriIli}. A rough estimate of $\beta$ can be made identifying $\beta^{1/2}$ with the normalized length of grain $\frac{\Delta x}{\lambda _{J}}$ \cite{nota3}. From the data of Ref.s \cite{Tafuri,floriana} is found $\lambda_J\sim 5$ $\mu$m which gives $\beta\simeq 0\allowbreak .04$. GBs faceting will give also a smaller $\beta$. 
  
By integrating Eq.(2) we find the phases for all junctions. Initially the phases of conventional junctions are set to zero and the phases of $\pi$-junction to $\pi $ or $-\pi $, which are the stable equilibrium points of the single junction potential.  This two possible choices correspond to two different sign of the spontaneous current circulating around $\pi$-loops. $\alpha=1/\sqrt{\beta_C}$ was set to $0.25$, which is within the interval proposed in \cite{Edward}. We do not use a field cooling process like in \cite{Deleo} because initial conditions naturally set out diamagnetic or paramagnetic solution like in the single loop. In absence of bias current, the system naturally sets in a static equilibrium solution (ground state \cite{Edward}) after few plasma periods. Then the local magnetization is evaluated by: 
\begin{equation}
m_{j}=\frac{\Phi _{tot,j}}{\Phi _{0}}-\frac{\Phi _{ext}}{\Phi _{0}}=\frac{%
\Delta \varphi _{j}}{2\pi }-f
\end{equation}
where $\Delta\varphi_j=\varphi_{j+1}-\varphi_{j}$ and the mean magnetization by: 
\begin{equation}
m=\frac{1}{N}\sum m_{j}=\frac{1}{N}\frac{\sum \Delta \varphi _{j}}{2\pi }-f=%
\frac{\Delta \varphi }{2\pi N}-f
\end{equation}
where $\Delta\varphi=\varphi_{N+1}-\varphi_1$.
In the absence of an external magnetic field the magnetization for a single localized $\pi$-loop in the array center (symmetric $0-\pi$ junction \cite{KirtleyMoler}) have the shape reported in Fig.2 topmost curves where the two spontaneous magnetization are shown for a $N=63$ loop array with $\beta =0.04$. The shape is very similar of "half-fluxon" obtained in the continuous approach \cite{Edward} due to relatively small $\beta$. In Fig. 2 the effect of the magnetic field increase on the spontaneous magnetizations is also shown. The magnetic field breaks the symmetry of two solutions: one is paramagnetic and the other diamagnetic. With the increase of the magnetic field the magnetization of the paramagnetic state is progressively reduced due to the screening diamagnetic currents that are generated at the boundary. The same currents add to the magnetization of the diamagnetic state giving a larger diamagnetic magnetization.

\begin{figure}[t]
\includegraphics{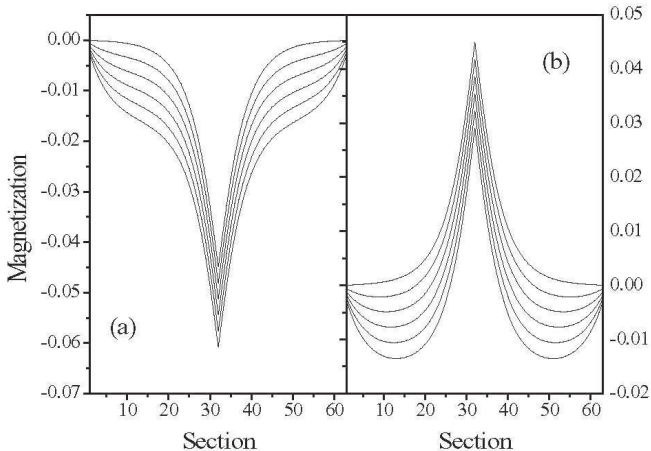}
\caption{Simulated magnetization of a $N=63$ Josephson junction array with a single $\pi$-loop  in the middle with $\beta_L=0.04$ and $\alpha=0.25$: (a) diamagnetic solution with progressively increasing magnetic field $\eta$ top to bottom $0$, $0.1$, $0.2$, $0.3$, $0.4$, $0.5$ (b) paramagnetic solution with progressively increasing magnetic field [same values of (a)].} 
\end{figure}

In Fig. 3 the mean magnetization for an array with a single $\pi$-loop is reported (circles). We note that magnetization of paramagnetic state is zero at a threshold field $\eta^*\simeq 0.29$. The linear decrease of mean is similar to that observed for (large $\beta$) single loops \cite{icem2002}. For the parameters of Fig. 2 the physical threshold field is $B^*\sim 38$ mG with the $\lambda_L$ given in Ref. \cite{Tafuri}. 

\begin{figure}
\includegraphics{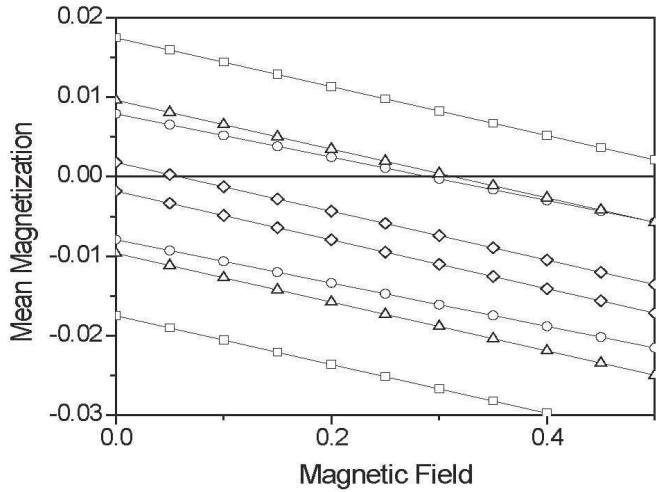}
\caption{Mean magnetization of both paramagnetic (upper curve) and diamagnetic (lower curve) solutions for Josephson junction mixed arrays. For all curves $\alpha=0.25$ and $\beta=0.04$: $\circ$ $N=63$ with a single $\pi$-loop; $\Diamond$ $N=255$ with 15 $\pi$-loops and one odd paramagnetic half flux quantum; $\triangle$ $N=255$ with 15 $\pi$-loops and $10$ paramagnetic half flux quanta; $\Box$ $N=255$ with 15 $\pi$-loops and $12$ paramagnetic half flux quanta.}
\end{figure}

In Fig.4a is reported the magnetization pattern in an array of $N=255$ loops with $15$ localized 
$\pi$-loops. According to Ref. \cite{Edward} flux quanta are sufficiently separated here to stay stable being the (minimum) length of conventional or $\pi$ sections $\Delta x/\lambda_J \simeq 4.64$.
The solution shows seven pairs positive-negative of half flux quanta plus an unpaired half flux
quantum. In Fig.4a the unpaired half flux quantum is positive, so solution is paramagnetic. An analogous diamagnetic solution exists when the unpaired half quantum is negative. Even $\pi$-loops configuration have zero spontaneous magnetization and are diamagnetic in small fields. Unpaired paramagnetic half flux quanta can be induced in the sample by a (moderate) field cooling process in small field, similar to \cite{Tafuri}. The behavior of the mean magnetization is reported Fig. 3 . Both the spontaneous magnetization and threshold field are very small in this case. With the above data we find $B^*\sim 7.6$ mG. In the same Fig.4 is also reported the case in which $10$ (Fig. 4b) and $12$ (Fig. 4c) $\pi$-loops have initial paramagnetic magnetization, which correspond to a stronger field cooling effect \cite{notaFC}. The corresponding mean magnetizations are again reported in Fig.3. The mean magnetization for $12$ paramagnetic $\pi$-loops becomes zero at $\eta^*\simeq 0.6$ which corresponds to $B^*\sim 80$ mG.

For the sake of clarity and brevity, the results shown above have been obtained in absence of disorder. Disorder has to be taken into account when we aim to describe high-T$_c$ materials and this will be subject of future investigations. Here we just observe that disorder can locally change the penetration length altering the section length $\Delta x/\lambda_J$ and/or permitting larger screening currents in the sample. Small $\Delta x/\lambda_J$ implies that "currentless" (constant phase) states can occur \cite{Mints,Edward} without spontaneous currents. These facts, together with the small values of the above threshold fields, imply that could be not surprising that also in moderate fields the state is diamagnetic \cite{Tafuri}. Therefore, the presence (or the absence) of spontaneous currents would be no more strictly correlated to paramagnetism. In \cite{TafuriIli} paramagnetism actually appears without measurable spontaneous currents in Scanning SQUID microscope images. 

\begin{figure}
\includegraphics{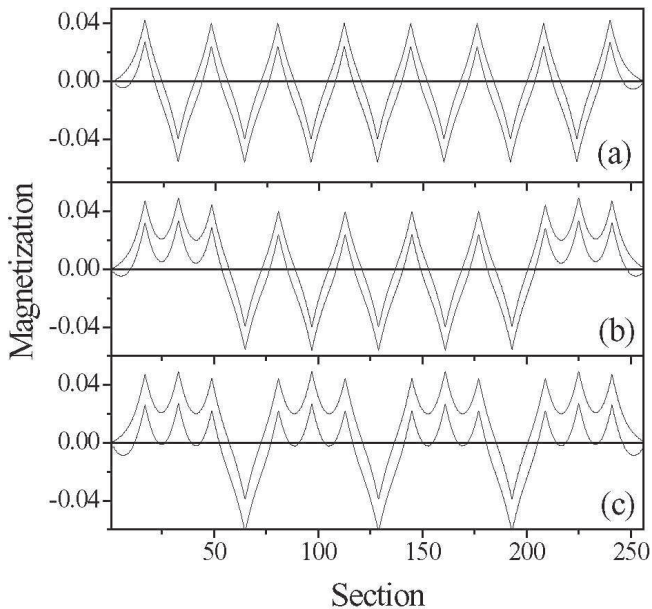}
\caption{Simulated magnetization of a $N=255$ Josephson junction 15 $\pi$-loop array with $\beta_L=0.04$ and $\alpha=0.25$: (a) solution with one unpaired paramagnetic half flux quantum, top curve $\eta=0$ bottom curve $\eta=0.1$; (b) solution with $10$ paramagnetic half flux quanta, top curve $\eta=0$ bottom curve $\eta=0.5$; (c) solution with $12$ paramagnetic half flux quanta, top curve $\eta=0$ bottom curve $\eta=0.7$.}
\end{figure}

In conclusion localized $\pi$-loops in GBs can show both spontaneous magnetization and paramagnetic behavior. For samples large with respect to the penetration depth, implying a low $\beta$ for each loop, paramagnetism exists in a relatively narrow region just near zero field. In absence of significant field cooling effects the energy difference between diamagnetic and paramagnetic fundamental state solutions can be very small so observation of paramagnetism can be difficult or strictly depending on the particular sample. Moreover in high-T$_c$ materials disorder can easily hinder the above picture. It is simpler to probe paramagnetic and diamagnetic states similar to that reported in Fig.2 for engineered systems of $\pi$-loops as recently reported in Ref. \cite{Hilgen03} for two 
dimensional systems. 
  
I warmly thank F. Tafuri, F. Lombardi, C. De Leo, P. Barbara and C. J. Lobb for useful discussions and suggestions. I must also thank E. Goldobin for having shown me his work. We acknowledge financial support from MIUR\ COFIN2000 project ''Dynamics and Thermodynamics of vortex structures in superconductive tunneling''.


\begin{thebibliography}{99}
\bibitem{notameta} It would be noted that for large $\beta$ loops Eq.(1) admit also higher order solutions typically referred as metastable states.
\bibitem{nota3} Using data on density current reported in Ref.s \cite{floriana,TafuriIli} the Josephson penetration length's are roughly similar to $\lambda_L$ which here is identified with the penetration length $\lambda_c$ given in Ref.\cite{Tafuri}.
\bibitem{notaFC} We have tested that this "field cooling at hand" is equivalent, except for
the location of "semifluxons", to take a random distribution of phases and cooling increasing $\beta$ to its final value in a given $f$. This procedure was used in larger 2d systems where it is not easy to choose an initial distribution of phases \cite{letter}.
\end{thebibliography}
\end{document}